\documentclass[journal]{IEEEtran}
\usepackage{epsfig}

\usepackage{cite}

\ifCLASSINFOpdf
\else
\fi
\usepackage{amsmath}
\usepackage{amssymb}
\usepackage{bm}
\usepackage{cases}
\usepackage{array}
\usepackage{url}

\begin{document}
%
\title{Efficient Numerical Modeling of the Magnetization Loss on a Helically Wound Superconducting Tape in a Ramped Magnetic Field}
%
%
%

\author{
Yoichi~Higashi~and~Yasunori~Mawatari
\thanks{Y. Higashi and Y. Mawatari are with National Institute of Advanced Industrial Science and Technology (AIST), Tsukuba, Ibaraki, 305-8568 Japan (e-mail: y.higashi@aist.go.jp).}
\thanks{Manuscript received April 3, 2019; revised July 22, 2019; accepted August 13, 2019.}
}

%
%

\markboth{Journal of \LaTeX\ Class Files,~Vol.~14, No.~8, August~2015}%
{Shell \MakeLowercase{\textit{et al.}}: Bare Demo of IEEEtran.cls for IEEE Journals}
%



\maketitle

\begin{abstract}
We investigate theoretically the dependence of magnetization loss of a helically wound superconducting tape  on the round core radius $R$ and the helical conductor pitch in a ramped magnetic field. 
Using the thin-sheet approximation, 
we identify the two-dimensional equation that describes Faraday's law of induction on a helical tape surface in the steady state. 
Based on the obtained basic equation, 
we simulate numerically the current streamlines and the power loss $P$ per unit tape length on a helical tape.
For $R \gtrsim w_0$ (where $w_0$ is the tape width), 
the simulated value of $P$ saturates close to the loss power $\sim(2/\pi)P_{\rm flat}$ (where $P_{\rm flat}$ is the loss power of a flat tape) for a loosely twisted tape.
This is verified quantitatively by evaluating power loss analytically in the thin-filament limit of $w_0/R\rightarrow 0$. 
For $R \lesssim w_0$, upon thinning the round core, 
the helically wound tape behaves more like a cylindrical superconductor as verified by the formula in the cylinder limit of $w_0/R\rightarrow 2\pi$, and 
$P$ decreases further from the value for a loosely twisted tape, reaching $\sim (2/\pi)^2 P_{\rm flat}$.
\end{abstract}


%
\IEEEpeerreviewmaketitle

\section{Introduction}
\IEEEPARstart{A}{} rare-earth barium copper oxide (ReBCO) YBa$_2$Cu$_3$O$_{7-\delta}$ coated conductor on round core (CORC) cable was proposed a decade ago \cite{laan2009} 
as a new concept for practical superconducting (SC) coated conductor cables. 
This has since been developed into a much thinner CORC {\it wire} with a round core diameter of a few millimeters \cite{laan2016,weiss2017} 
by using an alloy substrate with a thickness of as little as 30~$\mu$m. 
Flexible deformation along the former  has ever been demonstrated, although such deformation degrades the critical current to some extent \cite{weiss2017,souc2010,laan2011}. 
Furthermore, the current technology for fabricating CORC wires achieves the shortest conductor pitch (less than 10~mm \cite{weiss2017}) among all high-temperature superconductor (HTS) cable concepts.
Therefore, CORC wires show promise as practical SC wires for applications such as the high-field magnets used in particle accelerators and magnetic resonance imaging (MRI) machines.

To date, there have been relatively few reports on magnetization and transport alternating current (AC) loss measurements of CORC cables or wires without striation \cite{souc2010,majoros2014,terzioglu2017}. 
A helically wound tape structure (i.e., a continuous inclination of the tape) has been reported to decrease the magnetization loss 
because of the reduced average perpendicular field component penetrating the tape over a full pitch \cite{souc2010}.
However, other measurements have been reported showing higher AC loss in a helically wound tape in the fully penetrated state than in 
a flat tape. 
This fact is attributed to the anisotropy of the critical current density $J_{\rm c}$ with respect to the applied field \cite{majoros2014}; 
a helically wound tape has a higher $J_{\rm c}$ than that of a flat tape in a certain portion of the tape oriented parallel to the field. 
This also results in a sizable AC loss reduction relative to the loss in a flat tape at low field strength \cite{majoros2014}. 
For a single-layer CORC cable in the fully penetrated state, using more 
tapes in a single layer decreases the AC loss proportionally \cite{majoros2014}. 
Meanwhile, in the case of a multi-layered CORC cable in the fully penetrated state, 
the AC loss increases in proportion to the number of layers in the cable \cite{majoros2014}. 
An effect of a copper tube former on the magnetization and transport AC losses has been reported:
eddy current loss makes a large contribution to the total transport AC loss,
and the magnetic shielding reduces the magnetization AC loss at low field strength \cite{terzioglu2017}.

Regarding simulation studies, 
numerical simulations of AC losses based on the thin-sheet approximation have been carried out, the 
focus being the spiraled multi-layer coated conductors used for SC power transmission cables \cite{takeuchi2011,amemiya2013}. 
Another approach to determining the electromagnetic properties of a power transmission cable composed of helically wound SC tapes is based on electric-circuit theory using the Bean model \cite{tominaka2009}. 
A computational method was developed by transforming three-dimensional helicoidal structures into two-dimensional ones \cite{stenvall2013}. 
To simulate a CORC cable, the 
AC losses were simulated at first by a two-dimensional model neglecting the winding of the tapes \cite{solovyov2014}. 
Later, an improved simulation study of the magnetization losses in CORC cables was carried out using three-dimensional models based on the H formulation \cite{terzioglu2017,sheng2017}.
Simulations with three-dimensional models using the T-A formulation have also been conducted to investigate the AC losses in CORC cables, twisted stacked-tape conductor cables, and double coaxial cables with various transport currents and background magnetic fields \cite{fu2018}. Comprehensive 
simulations of the magnetization losses of CORC cables wound from the SC tapes with and without striation have been reported recently based on 
a three-dimensional model using the T-A formulation \cite{wang2019}.

However, an analytical approach to modeling the AC losses of a CORC wire is currently lacking, partly because of the rather complex wire structure, 
and therefore the theoretical limit of the AC loss reduction in a CORC wire remains elusive.
Herein, we develop an analytical method for modeling the magnetization loss of a helically wound SC tape. The 
magnetization loss is simulated to obtain properties such as its dependences on $R$ and $L_{\rm p}$ (see Fig.~\ref{fig1}), and
the simulation results are verified by the analytical formula in the limiting case of $w_0/R \rightarrow 2\pi$ or $0$. 
We consider the situation in which ReBCO tapes wound round an MRI magnet are exposed to a magnetic field that is ramped up to high strength \cite{yokoyama2017,yachida2017}.
The fully penetrated state is realized, resulting in the steady state in which the magnetic field profile on the tape is unchanged. 
Although the effects of tape stacking and $J_{\rm c}$ anisotropy on the AC losses are important issue \cite{majoros2014} for the quantitative evaluation, 
we ignore such effects and focus instead on the simplest model. 
The presented models and results represent a first step toward further advanced research and development of a high-field magnet wound from CORC wires.

\section{Model of a helically wound superconducting tape on a hollow cylinder}
\begin{figure}[tb]
\centering
\includegraphics[width=3.0in]{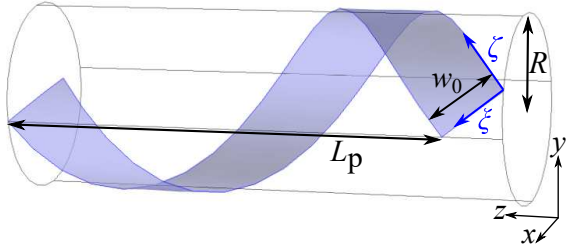}
\caption{Schematic of a helically wound superconducting (SC) tape of width $w_0$ and pitch $L_{\rm p}$ on a hollow cylinder of radius $R$. 
The axes $(\xi,\zeta)$ are on the tape surface.}
\label{fig1}
\end{figure}

We consider a helically wound SC tape on a hollow cylinder as shown in Fig.~\ref{fig1}. 
The tape width and thickness and the radius of the hollow cylinder are $w_0$, $d_0$, and $R$, respectively. 
The helically wound SC tape in the present simple model is assumed to be sufficiently thin ($d_0 \ll w_0$) that 
it can be approximated as a helical SC tape surface with infinitesimal thickness as depicted in Fig.~\ref{fig1}. 
The SC tape surface is described by the coordinates (see Appendix \ref{appendix_reduced-faraday})
\begin{equation}
\label{helicoid-coordinate}
\setlength{\nulldelimiterspace}{0pt}
\left\{
\begin{IEEEeqnarraybox}[\relax][c]{l's}
x=R \cos[ (k\zeta-\xi/R )/\sqrt{1+(kR)^2} ],\\
y=R \sin[(k\zeta-\xi/R)/\sqrt{1+(kR)^2}],\\
z=(\zeta+kR\xi)/\sqrt{1+(kR)^2},
\end{IEEEeqnarraybox}
\right.
\end{equation}
where $k=2\pi/L_{\rm p}$ and $L_{\rm p}$ is the pitch length of the helical conductor. 
The helically wound SC tape surface with a full pitch $L_{\rm p}$ corresponds to $-w_0/2\le\xi\le w_0/2$ and $0\le\zeta \le L_{\rm tape}$, 
with $L_{\rm tape}=L_{\rm p} \sqrt{1+(k R)^2}$ being the tape length corresponding to $L_{\rm p}$.
The $\xi$ and $\zeta$ axes are orthogonal to each other on the helical SC tape surface as shown in Fig.~\ref{fig1}.

\section{Response to an applied field on a helical superconducting tape}
\subsection{Reduced Faraday--Maxwell equation}
\begin{figure}[tb]
\centering
\includegraphics[width=1.3in]{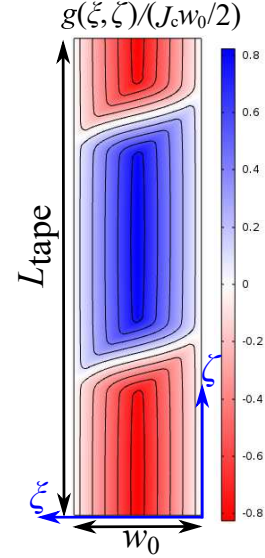}
\caption{Current streamlines and profile of the scalar function $g(\xi,\zeta)$ on a helically wound SC tape surface with $w_0=2$~mm and $L_{\rm p}=12$~mm 
for $\beta=3$~mT/s and $R=1.5$~mm. 
The tape length for one pitch of the helical conductor is $L_{\rm tape} \approx 15.26$~mm.
}
\label{fig2}
\end{figure}

We assume that the superconductor is in the steady state, which is plausible at high applied fields stronger than the full-flux-penetration field. 
We therefore neglect the contribution from the transport current because the magnetization losses are dominant over the transport losses at high field strength \cite{kajikawa2016}. We also ignore 
the magnetic field due to the screening current because that field is much weaker than the applied field. 
In an applied field $\bm{B}_{\rm a}(t)=B_{\rm a}(t)\hat{\bm{x}}$ that is ramped constantly at a rate $\beta={\rm d}B_{\rm a}(t)/{\rm d}t$, 
the Faraday--Maxwell equation in the steady state is 
\begin{equation}
\bm{\nabla}\times\bm{E}=-\frac{\partial \bm{B}}{\partial t} \approx -\frac{{\rm d} \bm{B}_{\rm a}(t)}{{\rm d} t} = -\beta \hat{\bm{x}},
\label{Faraday-maxwell-equation-steady}
\end{equation}
where $\bm{E}$ (resp.\ $\bm{B}$) is the electric (resp.\ magnetic) field and 
$\hat{\bm{x}}$ is the unit vector in the $x$ direction. 
With the thin-sheet approximation \cite{zhang2017},
we take account of only the response to the applied field component perpendicular to the tape surface.
Consequently, the Faraday--Maxwell equation reduces to the following two-dimensional equation on the tape surface \cite{higashi2019,higashi2018} (see Appendix \ref{appendix_reduced-faraday}):
\begin{equation}
\frac{\partial}{\partial \xi}\left(\rho_{\rm sc}\frac{\partial g}{\partial \xi} \right)+ \frac{\partial}{\partial \zeta}\left( \rho_{\rm sc}\frac{\partial g}{\partial \zeta} \right)=\beta \cos\left( \frac{k\zeta-\xi/R}{\sqrt{1+(kR)^2}}\right),
\label{reduced-FM-equation}
\end{equation}
where $g(\xi,\zeta)$ a scalar function defined on the SC tape surface. 
The contour lines of $g(\xi,\zeta)$ describe the current streamlines on the tape surface, as depicted by the solid lines in Fig.~\ref{fig2}.
The SC property is incorporated through the electric field ($\bm{E}$)--current density ($\bm{J}$) characteristics, namely $\bm{E}=\rho_{\rm sc}\bm{J}$, with
\begin{equation}
\rho_{\rm sc}(|\bm{J}|)=\frac{E_{\rm c}}{J_{\rm c}}\left( \frac{|\bm{J}|}{J_{\rm c}} \right)^{n-1}
\end{equation}
being the isotropic SC nonlinear resistivity. Herein, 
the electric field criterion is fixed to $E_{\rm c}=1$~$\mu$V/cm.

\subsection{Boundary conditions}
The specifications of the helically wound SC tape are summarized in Table~\ref{table1}. 
We simulate the current streamlines and the magnetization losses by solving Eq.~(\ref{reduced-FM-equation}) numerically using the commercial software COMSOL Multiphysics\textsuperscript{\textregistered} \cite{comsol}. 
The Dirichlet boundary condition $g(\xi=\pm w_0/2,\zeta)=0$ is imposed on $g(\xi,\zeta)$ along the long edges of the SC tape, 
whereas the periodic boundary condition $g(\xi,\zeta)=g(\xi,\zeta+L_{\rm tape})$ is imposed at the terminals of the SC tape with a full pitch. Herein, 
the field sweep rate is set to $\beta=3$~mT/s because the typical field sweep rate of an MRI magnet is of the order of milliteslas per second \cite{yokoyama2017,yachida2017}.

\section{Results and discussion}

\begin{table}
\caption{Specifications of helically wound SC tape}
\label{table}
\setlength{\tabcolsep}{3pt}
\begin{tabular}{|p{25pt}|p{106pt}|p{100pt}|}
\hline
Symbol& 
Quantity& 
Dimensions of helically wound SC tape \\
\hline
$w_0 $& tape width& 2~mm \\
$d_0 $&  tape thickness &2 $\mu$m\\
$L_{\rm p}$ & pitch length of helical conductor& 2--200~mm \\
$R$ & radius of hollow cylinder & 1/$\pi$--20~mm\\
\hline
&&Superconducting characteristics\\
\hline
$J_{\rm c}$& critical current density & 5$\times$10$^{10}$~A/m$^2$\\
$n$ & $n$-value & 29 \\
\hline
\end{tabular}
\label{table1}
\end{table}

\subsection{Current streamlines}
By solving Eq.~(\ref{reduced-FM-equation}) numerically, 
the current density vector on the tape surface can be calculated via
\begin{equation}
\bm{J}(\xi,\zeta)=J_\xi \hat{\bm{\xi}}+J_\zeta\hat{\bm{\zeta}},~~  J_\xi=-\frac{\partial g}{\partial \zeta},~~J_\zeta=\frac{\partial g}{\partial \xi},
\label{current-density}
\end{equation}
where $\hat{\bm{\xi}}$ (resp.\ $\hat{\bm{\zeta}}$) is the unit vector in the $\xi$ (resp.\ $\zeta$) direction.
The color scale in Fig.~\ref{fig2} shows $g(\xi,\zeta)/(J_{\rm c}w_0/2)$. The solid lines are the current streamlines, which 
are divided into two parts within a full pitch because of the helical winding, 
thereby reducing the effective tape length to $L_{\rm tape}/2$.

\subsection{Magnetization loss}
\begin{figure}[tb]
\centering
\includegraphics[width=3.5in]{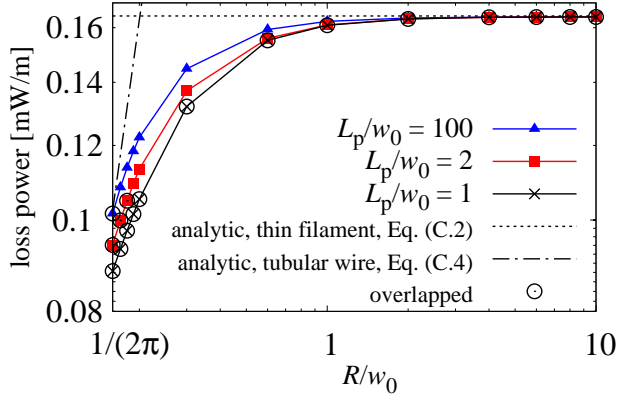}
\caption{Dependence of loss power per unit tape length on cylinder radius (plotted double logarithmically). 
The circles indicate where the SC tape overlaps.
}
\label{fig3}
\end{figure}

\begin{figure}[tb]
\centering
\includegraphics[width=3.5in]{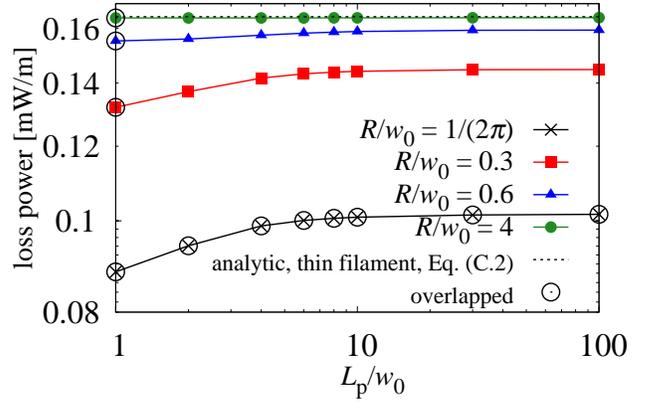}
\caption{Dependence of loss power per unit tape length on the pitch length of the helical conductor (plotted double logarithmically). 
The circles indicate where the SC tape overlaps.
}
\label{fig4}
\end{figure}

\begin{figure}[tb]
\centering
\includegraphics[width=3.5in]{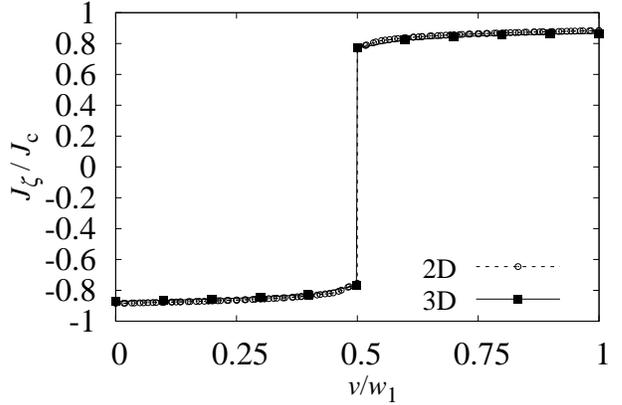}
\caption{Current density profile on a helical tape in the two and three dimensional models.
The pitch length and the radius of the hollow cylinder are set to $L_{\rm p}=10 w_0$ and $R=w_0$, respectively.
The other parameters are common.
}
\label{fig:current}
\end{figure}

The loss power $P$ per unit tape length is evaluated by (see Appendix \ref{appendix_coordinate-transformation})
\begin{equation}
P=\frac{d_0}{L_{\rm tape}}\int {\rm d}\xi\int{\rm d}\zeta \rho_{\rm sc}(\xi,\zeta)\left[ \left( \frac{\partial g}{\partial \xi}\right)^2+\left( \frac{\partial g}{\partial \zeta}\right)^2\right].
\end{equation}

Figure~\ref{fig3} shows the double logarithmic plot of the loss power per unit tape length versus $R$ for different values of $L_{\rm p}$.
At $R/w_0=1/2\pi$, the helical SC tape overlaps with itself irrespective of $L_{\rm p}$,
and for $L_{\rm p}/w_0=1$, the SC tape overlaps irrespective of $R$.
The circles in Fig.~\ref{fig3} indicate 
where the tape overlaps. 

For $R\gtrsim w_0$, 
the loss power saturates close to the analytically evaluated value (the dashed line in Fig.~\ref{fig3}) in the thin-filament limit of $w_0/R\rightarrow 0$ via Eq.~(\ref{loss_thin-filament}). 
For $n=29$, Eq.~(\ref{loss_thin-filament}) yields $P(w_0/R\rightarrow 0)\approx0.165$~mW/m, 
which is close to $(2/\pi)P_{\rm flat}$ (where $P_{\rm flat}$ is the loss power of the flat tape) in the Bean limit of $n\rightarrow \infty$ 
as evidenced by a recent magnetization measurement \cite{myers2019} [see also Eq.~(\ref{loss_thin-filament})].

Meanwhile, for $R\lesssim w_0$, with decreasing $R$ the loss power becomes much 
lower than that in the limit of $w_0/R\rightarrow 0$ (the dashed line in Fig.~\ref{fig3}). 
Upon decreasing $R/w_0$, the helically wound tape behaves more like a tubular wire, as discussed in Ref.~\cite{majoros2014}. 
Indeed, the loss power for $L_{\rm p}/w_0=100$ at $R/w_0=1/2\pi$ shows quantitatively good agreement with the analytically evaluated loss power (the dashed-dotted line in Fig.~\ref{fig3}) 
for a tubular wire via Eq.~(\ref{loss_cylinder}), 
although there is some deviation from the theoretical value for a tubular wire for $L_{\rm p}/w_0=1$ and 2 because of the short pitch. 
The slope of the dashed-dotted line is two in the Bean limit 
because the loss power for a tubular superconductor increases with $R^2$ as in Eq.~(\ref{loss_flat-tape_tube}). 
The initial $R$ slope for each value of $L_{\rm p}/w_0$ is less than two, although it approaches two with increasing $L_{\rm p}/w_0$. 
We ascribe the initial slope being less than two to the finite pitch effect specific to the helically wound tape.

Figure~\ref{fig4} shows the dependence of the loss power per unit tape length on $L_{\rm p}$ in double logarithm form for different values of $R$. 
The loss power shows weak $L_{\rm p}$ dependence for $L_{\rm p}/w_0 \lesssim 10$. 
At $R/w_0=4$, the loss power coincides with that evaluated via Eq.~(\ref{loss_thin-filament}) in the thin-filament limit (the dashed line in Fig.~\ref{fig4}) independent of $L_{\rm p}$. 
Meanwhile, for $R/w_0 <4$, 
the loss power becomes lower than the theoretical value in the thin-filament limit 
and becomes even lower with decreasing $R$ as already mentioned in Fig.~\ref{fig3}. 
The circles in Fig.~\ref{fig4} indicate where the tape overlaps.

Figure~\ref{fig:current} shows the profiles of the current density $J_{\zeta}(v,\zeta=0.5L_{\rm p})$ in the direction of the long dimension of a helical tape in two and three dimensional models.
The profile of $J_{\zeta}(v,\zeta=0.5L_{\rm p})$ on a helical tape shows a quantitative agreement between two and three dimensional models.
Here, the axes $v$ and $\zeta$ are defined via $(x,y,z)=(R\cos k(\zeta-v),R\sin k(\zeta-v),\zeta)$.

Next, we explain
physically why the loss power decreases for $R \lesssim w_0$. 
The length scale for the penetration of magnetic flux vortices decreases as $R$ decreases, thereby 
reducing the AC losses \cite{mawatari2011}.
In the present study, we consider the steady state wherein the applied field should far exceed the full-flux-penetration field. 
We therefore consider that it is 
the reduced flux-penetration length scale, rather than the screening of the applied magnetic field, that reduces the loss power.

Next, we evaluate approximately the theoretically expected maximum reduction rate of the loss power upon reducing $R$. 
We evaluate the ratio of the loss power for $R/w_0=1/2\pi$ to that for $R/w_0\gg 1$ (i.e., $w_0/R\rightarrow 0$) by using Eqs.~(\ref{loss_thin-filament}) and (\ref{loss_cylinder}):
\begin{equation}
\frac{P(w_0/R\rightarrow 2\pi)}{P(w_0/R\rightarrow 0)}=\frac{2+1/n}{\pi^{1+1/n}}\rightarrow \frac{2}{\pi}~~(n\rightarrow \infty).
\label{reduction-rate}
\end{equation}
This suggests that by thinning the wire radius down to $R\ll w_0$, the loss power of the helical tape becomes smaller than $P_{\rm flat}$ by a factor of $(2/\pi)^2\approx 0.405$ 
in accordance with a previous study \cite{mawatari2011}. Our numerical results 
confirm the reduction rate of the loss power. 
At $L_{\rm p}/w_0=100$, the ratio of the loss power for $R/w_0=1/2\pi$ to that for $R/w_0=4$ is 
$P(R/w_0=1/2\pi)/P(R/w_0=4)\approx 0.619$, which agrees well with $(2+1/n)/\pi^{1+1/n}\approx 0.623$ [Eq.~(\ref{reduction-rate})] for $n=29$.

\begin{figure}[tb]
\centering
\includegraphics[width=3.1in]{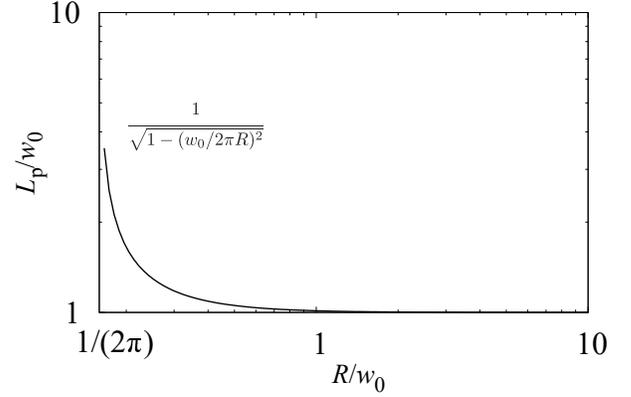}
\caption{Critical line on which the helically wound SC tape begins to overlap.
}
\label{fig5}
\end{figure}

Next, we consider the condition that must be met for the helically wound SC tape not to overlap. 
This is determined by $L_{\rm p}>w_0/\cos \alpha$ (see Appendix \ref{appendix_reduced-faraday}.1).
By solving the condition with respect to $L_{\rm p}$, we obtain the condition that $L_{\rm p}>w_0/\sqrt{1-(w_0/2\pi R)^2}$ (see Fig.~\ref{fig5}).

\subsection{CORC wire with multiple tapes}
We discuss the effect of multiple tapes on magnetization losses in a single layer CORC wire.
The response to an applied field is sufficiently large at high fields far above a full flux penetration field $B_{\rm p}$ that the interaction between SC tapes is relatively small and negligible,
although the magnetic field shielding by neighboring tapes should be important to model a CORC wire wound by several tapes at low fields \cite{majoros2014,wang2019}.
The magnetic field due to the shielding current is of the order of $\mu_0 J_{\rm c}d_0$ ($\approx 0.126$ T for the parameters adopted in this paper).
It is sufficiently small comparing to the background field.
Thus, we can ignore the contribution from a shielding current and treat total losses as a simple sum of the magnetization loss on an individual helical tape.

Our assumption should be plausible when the background field strength is as high as a few Tesla or at least above $B_{\rm p}$.       
$B_{\rm p}$ for a single helical tape is approximately evaluated by the formula for a flat tape, $B_{\rm p}=(\mu_0 J_{\rm c}d_0/\pi)[1 + \ln(w_0/d_0)] \approx 0.316$ T \cite{brandt1996},
where $\mu_0$ is the magnetic permeability of a vacuum.

\section{Summary}
We investigated theoretically the magnetization loss of a helically wound SC tape on a hollow cylinder in the steady state under a ramped field. 
For $R\gtrsim w_0$, the loss power saturates close to the value expected for a loosely twisted SC tape. 
Meanwhile, for $R\lesssim w_0$, it becomes lower than that value. 
Compared with the loss power $P_{\rm flat}$ for a flat SC tape, 
the loss power for a helical SC tape is roughly that for a loosely twisted tape ($\sim 2/\pi$ times lower than $P_{\rm flat}$) for $R\gtrsim w_0$. 
Further reduction of the loss power value is expected by thinning the round core diameter down to $R\lesssim w_0$. 
For a thin wire with a radius of a few millimeters, 
the loss power per unit tape length is almost independent of $L_{\rm p}$.

\section*{Acknowledgment}
We would like to thank Y.~Yoshida (AIST) for providing us with the properties and dimensions of commercial superconducting CORC\textsuperscript{\textregistered} wires. 
We also thank R.~Toyomoto (Kyoto Univ.) and N.~Amemiya (Kyoto Univ.) for sharing unpublished AC-loss measurement and simulation data with us. 
The present work is based on the results obtained from a project commissioned by the New Energy and Industrial Technology Development Organization (NEDO).

\appendices
\renewcommand{\thesectiondis}[2]{\Alph{section}:}
%
\section{Derivation of reduced Faraday--Maxwell equation}
\label{appendix_reduced-faraday}
\renewcommand{\theequation}{A.\arabic{equation}}
\setcounter{equation}{0}

\renewcommand{\thesubsectiondis}[0]{\Alph{section}.\arabic{subsection}.}
\setcounter{subsection}{0}
\subsection{Model of helical tape}
\label{appendix_model-helical}
We use the following helical coordinates to model the helically wound SC tape:
\begin{equation}
\label{helix-coordinates}
\setlength{\nulldelimiterspace}{0pt}
\left\{
\begin{IEEEeqnarraybox}[\relax][c]{l's}
x=u \cos\theta , \\
y=u \sin\theta,\\
z=\theta/k+v,
\end{IEEEeqnarraybox}
\right.
\end{equation}
where $k=2\pi/L_{\rm p}$ and $L_{\rm p}$ is the pitch length of the helical conductor. 
The tape surface for one pitch length of the helical conductor corresponds to $u=R$, $0\le v \le w_0/\cos\alpha$, and $0\le\theta\le2\pi$, where 
$R$ is the radius of the hollow cylinder, $w_0$ is the tape width, and $\alpha$ is the lay angle defined by $\cos\alpha=kR/\sqrt{1+(kR)^2}$.

In general, 
the current density vector is described in terms of the current vector potential $\bm{T}$ as $\bm{J}=\bm{\nabla}\times \bm{T}$ so as to satisfy $\bm{\nabla}\cdot\bm{J}=0$. 
In the thin-sheet approximation, we regard the SC tape as a strip with infinitesimal thickness, 
and therefore the electric current flows on the tape surface at $u=R$. 
In this case, the current vector potential can be expressed in terms of the tape-normal unit vector $\bm{\nabla}u$ as $\bm{T}=\bm{\nabla}\times[g(\theta,v)\bm{\nabla}u]$. 
Here, the scalar function $g(\theta,v)$ describes the current streamlines on the tape surface. 
The current density on the tape surface at $u=R$ is calculated as
\begin{equation}
\bm{J}=\bm{\nabla}g(\theta,v)\times\bm{\nabla}u=J_v\hat{\bm{v}}+J_\theta\hat{\bm{\theta}},
\label{current-density-u-v-theta}
\end{equation}
where $\hat{\bm{v}}$ and $\hat{\bm{\theta}}$ are the unit vectors in the $v$ and $\theta$ directions, respectively, and
\begin{equation}
J_v=-\frac{1}{R}\frac{\partial g}{\partial \theta},~~~J_\theta=\frac{\sqrt{1+(kR)^2}}{kR}\frac{\partial g}{\partial v}.
\end{equation}

\subsection{Reduced Faraday--Maxwell equation on a helical tape}
The electric field in the helical coordinates (\ref{helix-coordinates}) is $\bm{E}=E_u\hat{\bm{u}}+E_v\hat{\bm{v}}+E_\theta\hat{\bm{\theta}}$ with
\begin{align}
\hat{\bm{u}}&=\bm{\nabla}u=\hat{\bm{x}}\cos\theta+\hat{\bm{y}}\sin\theta,\label{u-hat}\\
\hat{\bm{v}}&=\bm{\nabla}\theta/k+\bm{\nabla}v=\hat{\bm{z}},\label{v-hat}\\
\hat{\bm{\theta}}&=(\sqrt{1+(ku)^2}/k)\bm{\nabla}\theta+\bm{\nabla}v/\sqrt{1+(ku)^2}\nonumber \\
&=ku(-\hat{\bm{x}}\sin\theta+\hat{\bm{y}}\cos\theta+\hat{\bm{z}}/ku)/\sqrt{1+(ku)^2} \label{theta-hat}.
\end{align}
Then, the electric field is expressed as 
$\bm{E}=E_u\bm{\nabla}u+\tilde{E}_v\bm{\nabla}v+\tilde{E}_\theta\bm{\nabla}\theta$, 
where $\tilde{E}_v=E_v+E_\theta/\sqrt{1+(ku)^2}$ 
and $\tilde{E}_\theta =E_v/k+E_\theta\sqrt{1+(ku)^2}/k$. 
Using the relations $\bm{\nabla}u\times\bm{\nabla}v=-(\sqrt{1+(ku)^2}/ku)\hat{\bm{\theta}}$, $\bm{\nabla}v\times\bm{\nabla}\theta=-\hat{\bm{u}}/u$, and 
$\bm{\nabla}\theta \times \bm{\nabla}u=-\hat{\bm{v}}/u$, 
we obtain 
\begin{align}
\bm{\nabla}\times\bm{E}
&=-\left( \frac{\partial \tilde{E}_v}{\partial u}-\frac{\partial E_u}{\partial v} \right) \frac{\sqrt{1+(ku)^2}}{ku}\hat{\bm{\theta}} \nonumber\\
&-\left( \frac{\partial\tilde{E}_\theta}{\partial v}-\frac{\partial \tilde{E}_v}{\partial \theta}\right)\frac{\hat{\bm{u}}}{u}-\left(\frac{\partial E_u}{\partial \theta}-\frac{\partial \tilde{E}_\theta}{\partial u} \right)\frac{\hat{\bm{v}}}{u}.
\label{faraday-induction-polar}
\end{align}
Because we adopt the thin-sheet approximation, we may consider the response to the perpendicular field component alone. 
Therefore, projecting Eq.~(\ref{Faraday-maxwell-equation-steady}) with eq.~(\ref{faraday-induction-polar}) onto the unit vector normal to the tape surface $\hat{\bm{u}}$ gives 
\begin{align}
\frac{\partial}{\partial v}&\left( E_v/ku+\sqrt{1+(ku)^2}E_\theta/ku\right)\nonumber\\
&-\frac{1}{u}\frac{\partial}{\partial \theta}\left( E_v+E_\theta/\sqrt{1+(ku)^2}\right)=\beta \cos\theta
\label{Faraday-E-helical}
\end{align}
by using $\hat{\bm{u}}\cdot \hat{\bm{\theta}}=0$ and $\hat{\bm{u}}\cdot \hat{\bm{v}}=0$. 
Combining Eq.~(\ref{Faraday-E-helical}) with $(E_\theta,E_v)=\rho_{\rm sc} (J_\theta,J_v)$, the Faraday--Maxwell equation on the tape surface $u=R$ is reduced to 
\begin{align}
\frac{\partial }{\partial v}&\left[ \rho_{\rm sc} \frac{1+(kR)^2}{(kR)^2}\frac{\partial g}{\partial v}-\frac{\rho_{\rm sc}}{kR}\frac{1}{R}\frac{\partial g}{\partial \theta} \right]\nonumber \\
&+\frac{1}{R}\frac{\partial}{\partial \theta}\left[\frac{\rho_{\rm sc}}{R}\frac{\partial g}{\partial \theta}-\frac{\rho_{\rm sc}}{kR}\frac{\partial g}{\partial v}  \right]=\beta\cos\theta.
\label{reduced-Faraday}
\end{align}

Although the derivation of the reduced Faraday--Maxwell equation (\ref{reduced-Faraday}) is clear in the coordinates of Eq.~(\ref{helix-coordinates}), 
Eq.~(\ref{reduced-Faraday}) includes the divergent terms in the loosely wound limit of $kR\rightarrow 0$ as is obvious from geometric considerations. We 
therefore work in the coordinates in which the spatial axes $(\xi,\zeta)$ are orthogonal on the tape surface. 
An arbitrary position on the tape surface in the $(\xi,\zeta)$ coordinates is expressed in terms of $(v,\theta)$ as
\begin{equation}
\xi=v  \cos\alpha,~~\zeta=\sqrt{(R\theta)^2+(\theta/k)^2}+v\sin\alpha,
\end{equation}
from which we have that
\begin{equation}
v=\frac{\sqrt{1+(kR)^2}}{kR}\xi,~~\theta=\frac{k\zeta-\xi/R}{\sqrt{1+(kR)^2}}.
\label{v-theta}
\end{equation}
Thus, on the tape surface $u=R$, from Eqs.~(\ref{helix-coordinates}) and (\ref{v-theta}), we obtain Eq.~(\ref{helicoid-coordinate}).

For $\bm{r}=x\hat{\bm{x}}+y\hat{\bm{y}}+z\hat{\bm{z}}$, we have that 
\begin{align}
\hat{\bm{\xi}}&=\partial \bm{r}/\partial \xi=-\hat{\bm{\theta}}/ku+(\sqrt{1+(ku)^2}/ku)\hat{\bm{v}},\label{xi-hat}\\
\hat{\bm{\zeta}}&=\partial \bm{r}/\partial \zeta=\hat{\bm{\theta}}\label{zeta-hat}.
\end{align}
Thus, we obtain 
\begin{eqnarray}
\hat{\bm{v}}=(\hat{\bm{\zeta}}+ku\hat{\bm{\xi}})/\sqrt{1+(ku)^2},~~\hat{\bm{\theta}}=\hat{\bm{\zeta}}.
\label{v-theta-hat}
\end{eqnarray}
As for the reduced Faraday--Maxwell equation, substituting the expressions
\begin{eqnarray}
\frac{1}{R}\frac{\partial g}{\partial \theta}=\frac{1}{R}\left( \frac{\partial g}{\partial \xi}\frac{\partial \xi}{\partial \theta}+\frac{\partial g}{\partial \zeta}\frac{\partial \zeta}{\partial \theta}\right)=\frac{\partial g}{\partial \zeta}\frac{1}{\cos\alpha},\\
\frac{\partial g}{\partial v}=\frac{\partial g}{\partial \xi}\frac{\partial \xi}{\partial v}+\frac{\partial g}{\partial \zeta}\frac{\partial \zeta}{\partial v}=\frac{\partial g}{\partial \xi}\cos\alpha+\frac{\partial g}{\partial \zeta}\sin\alpha
\label{g-derivative}
\end{eqnarray}
into Eq.~(\ref{reduced-Faraday}) yields Eq.~(\ref{reduced-FM-equation}). 
Furthermore, by substituting Eqs.~(\ref{v-theta-hat})--(\ref{g-derivative}) into Eq.~(\ref{current-density-u-v-theta}), 
we have the screening current density Eq.~(\ref{current-density}) on the tape surface $u=R$.

\section{Vectors in $(\xi,u,\zeta)$ coordinates}
\renewcommand{\theequation}{B.\arabic{equation}}
\setcounter{equation}{0}
\label{appendix_coordinate-transformation}
First, by using Eq.~(\ref{v-theta-hat}), 
the coordinate transformation from the $(\theta,u,v)$ coordinates to the $(\xi,u,\zeta)$ ones can be achieved for an arbitrary vector $\bm{O}=(O_x,O_y,O_z)$. 
Second, in the same way, $\bm{O}$ in the $(\xi,u,\zeta)$ coordinates is transformed to that in the $(x,y,z)$ ones by using Eqs.~(\ref{u-hat})--(\ref{theta-hat}), (\ref{xi-hat}), and (\ref{zeta-hat}):
$
\bm{O}=O_u\hat{\bm{u}}+O_v\hat{\bm{v}}+O_\theta\hat{\bm{\theta}}=O_u\hat{\bm{u}}+O_\xi\hat{\bm{\xi}}+O_\zeta\hat{\bm{\zeta}}=O_x\hat{\bm{x}}+O_y\hat{\bm{y}}+O_z\hat{\bm{z}},
$
where $O_\xi\equiv kuO_v/\sqrt{1+(ku)^2}$ and $O_\zeta\equiv O_\theta+O_v/\sqrt{1+(ku)^2}$. 
In the case of the thin-sheet approximation, we have on the tape surface $u=R$ that
\begin{equation}
\label{vector-thin-film-approx}
\setlength{\nulldelimiterspace}{0pt}
\left\{
\begin{IEEEeqnarraybox}[\relax][c]{l's}
O_x\approx(O_\xi \sin\alpha-O_\zeta \cos\alpha)\sin\theta, \\
O_y\approx-(O_\xi \sin \alpha-O_\zeta \cos \alpha)\cos\theta, \\
O_z\approx (O_\xi \cos\alpha +O_\zeta \sin\alpha).
\end{IEEEeqnarraybox}
\right.
\end{equation}
Herein, note that $O_u$ vanishes in the thin-sheet approximation when $\bm{O}$ refers to the current density or the electric field, neither of which has an out-of-plane component. 
Thus we have from Eq.~(\ref{vector-thin-film-approx}) that $O^2_x+O^2_y+O^2_z=O^2_\xi+O^2_\zeta$. 
Note also that the Jacobian $\partial(x,y,z)/\partial(\xi,u,\zeta)=1$.

\section{Magnetization loss in thin-filament or cylinder limit}
\renewcommand{\theequation}{C.\arabic{equation}}
\setcounter{equation}{0}
\label{appendix_gapped-cylinder}
\subsection{Thin-filament limit of $w_0/R \rightarrow 0$}
\label{appendix_thin-filament-limit}
\begin{figure}[tb]
\centering
\includegraphics[width=3.5in]{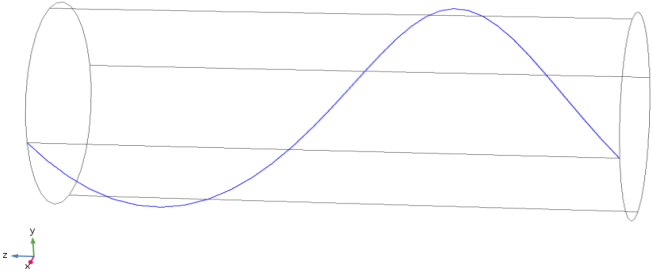}
\caption{Schematic of a helical tape in the thin-filament limit of $w_0/R \rightarrow 0$.}
\label{fig6}
\end{figure}
In the thin-filament limit, we may take the limit of $\xi/R\rightarrow 0$ while keeping $kR$ finite (see Fig.~\ref{fig6}). 
In this limit, the scalar function $g(\xi,\zeta)$ is almost uniform in the $\zeta$ direction, 
and therefore the second term on the left-hand side (l.h.s.) of Eq.~(\ref{reduced-FM-equation}) can be neglected because $\partial g/\partial \zeta \approx 0$. 
Thus, we obtain the equation for the scalar function:
\begin{equation}
\frac{\partial}{\partial \xi}\left( \rho_{\rm sc}\frac{\partial g}{\partial \xi} \right) \approx \beta\cos\phi(\zeta),
\label{eq_thin-filament}
\end{equation}
with $\phi(\zeta)=k\zeta/\sqrt{1+(kR)^2}$. 
Equation~(\ref{eq_thin-filament}) has been solved previously with the Dirichlet boundary condition $g(\xi=\pm w_0/2,\zeta)=0$ \cite{higashi2019}. 
The loss power per unit tape length coincides with that for a twisted tape \cite{higashi2019} and is readily evaluated as
\begin{align}
P\left({w_0\over R}\rightarrow 0\right)
&=\frac{B(\frac{2n+1}{2n},\frac{1}{2})}{\pi}\left( \frac{\beta w_0}{2E_{\rm c}} \right)^{1\over n}\frac{J_{\rm c} d_0 w^2_0 \beta}{2(2+1/n)}
\label{loss_thin-filament}\\
&\to (2/\pi)J_{\rm c}d_0w^2_0 \beta/4,~~~(n\to \infty), \nonumber
\end{align}
where $B(p,q)=B(q,p)=2\int_0^{\pi/2}{\rm d}\theta \cos^{2p-1} \theta \sin^{2q-1} \theta$ is the beta function and $p$ and $q$ are positive real numbers.

\subsection{Cylinder limit of $w_0/R \rightarrow 2\pi$}
\begin{figure}[tb]
\centering
\includegraphics[width=3.5in]{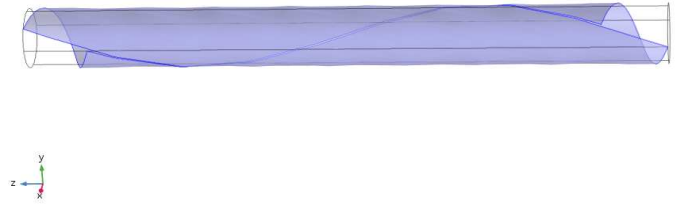}
\caption{Schematic of a helical tape in the limit of $w_0/R \rightarrow 2\pi$. By taking the limit of $kR \rightarrow 0$ as well, a helical tape approaches to a gapped cylindrical superconductor.}
\label{fig7}
\end{figure}
In the loosely wound limit of $L_{\rm p} \rightarrow \infty$ (i.e., $kR\rightarrow 0$), 
$g(\xi,\zeta)$ can be viewed as uniform in the $\zeta$ direction, 
and thus the second term on the l.h.s.\ of Eq.~(\ref{reduced-FM-equation}) may be neglected. 
We note that $\theta=(k\zeta-\xi/R)/\sqrt{1+(kR)^2}$ on the right-hand side of Eq.~(\ref{reduced-FM-equation}) should be kept over a full pitch to take account of the helical winding effect of the tape. 
However, we encounter severe difficulty in evaluating the loss power analytically in this case. 
Instead, we take the cylinder limit of $w_0/R\rightarrow 2\pi$ in addition to the limit of $kR \rightarrow 0$ to understand the dependence of the loss power on $R$ (see Fig.~\ref{fig7}).
By taking the limit of $kR\rightarrow 0$, we obtain 
\begin{equation}
\frac{\partial}{ \partial \xi} \left( \rho_{\rm sc} \frac{\partial g}{\partial \xi}\right) \approx \beta \cos(-\xi/R).
\label{eq_gapped-cylinder}
\end{equation}
In this case, we can neglect the $kR\zeta$ term because it is not necessary to keep $\theta=(k\zeta-\xi/R)/\sqrt{1+(kR)^2}$ over a full pitch $L_{\rm p}$ in the limit of $w_0/R \rightarrow 2\pi$. 
Equation~(\ref{eq_gapped-cylinder}) lacks the information about the helical structure of the tape in the $\zeta$ direction 
and thus describes Faraday's law of induction in the steady state for an infinitely long gapped cylindrical superconductor. 
The loss power of a tubular superconductor per unit tape length is readily evaluated in the limit of $w_0/R \rightarrow 2\pi$ as
\begin{equation}
P\left({w_0\over R}\rightarrow 2\pi\right)
=
2B\left({1\over2},{2n+1 \over 2n}\right)  J_{\rm c}d_0R^2\beta \left( {\beta R\over E_{\rm c}} \right)^{1/n}.
\label{loss_cylinder}
\end{equation}
Meanwhile, in the Bean limit of $n\rightarrow \infty$, the loss power of a gapped cylindrical superconductor per unit tape length is reduced to
\begin{align}
P_{\rm gapped}&
= 2J_{\rm c}d_0 \beta R^2\left[ 1-\cos(w_0/2R) \right]  \nonumber \\
&
\left\{
\begin{IEEEeqnarraybox}[\relax][c]{l's}
\approx J_{\rm c}d_0 w^2_0 \beta/4~~(w_0/R\ll 1),\\
\rightarrow 4J_{\rm c}d_0R^2\beta~~(w_0/R \rightarrow 2\pi),
\end{IEEEeqnarraybox}
\right.
\label{loss_flat-tape_tube}
\end{align}
which coincide with the loss power for a flat tape in a swept field \cite{brandt1996,higashi2019} and a tubular superconductor (the $R$ dependence is in accord with Ref.~\cite{mawatari2011}), respectively.

\ifCLASSOPTIONcaptionsoff
  \newpage
\fi

\bibliographystyle{IEEEtran}
\bibliography{IEEEabrv,IEEEexample}

\begin{thebibliography}{10}
\providecommand{\url}[1]{#1}
\csname url@samestyle\endcsname
\providecommand{\newblock}{\relax}
\providecommand{\bibinfo}[2]{#2}
\providecommand{\BIBentrySTDinterwordspacing}{\spaceskip=0pt\relax}
\providecommand{\BIBentryALTinterwordstretchfactor}{4}
\providecommand{\BIBentryALTinterwordspacing}{\spaceskip=\fontdimen2\font plus
\BIBentryALTinterwordstretchfactor\fontdimen3\font minus
  \fontdimen4\font\relax}
\providecommand{\BIBforeignlanguage}[2]{{%
\expandafter\ifx\csname l@#1\endcsname\relax
\typeout{** WARNING: IEEEtran.bst: No hyphenation pattern has been}%
\typeout{** loaded for the language `#1'. Using the pattern for}%
\typeout{** the default language instead.}%
\else
\language=\csname l@#1\endcsname
\fi
#2}}
\providecommand{\BIBdecl}{\relax}
\BIBdecl

\bibitem{laan2009}
D.~C. van~der Laan, ``{YBa}$_2${Cu}$_3${O}$_{7-\delta}$ coated conductor
  cabling for low ac-loss and high-field magnet applications,''
  \emph{Supercond. Sci. Technol.}, vol.~22, no.~6, p. 065013, 2009.

\bibitem{laan2016}
D.~C. van~der Laan, J.~D. Weiss, P.~Noyes, U.~P. Trociewitz, A.~Godeke,
  D.~Abraimov, and D.~C. Larbalestier, ``Record current density of 344
  {A}mm$^{-2}$ at 4.2 {K} and 17 {T} in {CORC}\textsuperscript{\textregistered}
  accelerator magnet cables,'' \emph{Supercond. Sci. Technol.}, vol.~29, no.~5,
  p. 055009, 2016.

\bibitem{weiss2017}
J.~D. Weiss, T.~Mulder, H.~J. ten Kate, and D.~C. van~der Laan, ``Introduction
  of {CORC}\textsuperscript{\textregistered} wires: highly flexible, round
  high-temperature superconducting wires for magnet and power transmission
  applications,'' \emph{Supercond. Sci. Technol.}, vol.~30, no.~1, p. 014002,
  2016.

\bibitem{souc2010}
J.~{\v{S}}ouc, M.~Vojen{\v{c}}iak, and F.~G\"om\"ory, ``Experimentally
  determined transport and magnetization ac losses of small cable models
  constructed from {YBCO} coated conductors,'' \emph{Supercond. Sci. Technol.},
  vol.~23, no.~4, p. 045029, 2010.

\bibitem{laan2011}
D.~C. van~der Laan, X.~F. Lu, and L.~F. Goodrich, ``Compact
  {G}d{B}a$_2${C}u$_3${O}$_{7-\delta}$ coated conductor cables for electric
  power transmission and magnet applications,'' \emph{Supercond. Sci.
  Technol.}, vol.~24, no.~4, p. 042001, 2011.

\bibitem{majoros2014}
M.~Majoros, M.~D. Sumption, E.~W. Collings, and D.~C. van~der Laan,
  ``Magnetization losses in superconducting {YBCO} conductor-on-round-core
  ({CORC}) cables,'' \emph{Supercond. Sci. Technol.}, vol.~27, no.~12, p.
  125008, 2014.

\bibitem{terzioglu2017}
R.~Terzio{\u{g}}lu, M.~Vojen{\v{c}}iak, J.~Sheng, F.~G\"om\"ory, T.~F.
  {\c{C}}avu{\c{s}}, and {\.{I}}.~Belenli, ``{AC} loss characteristics of
  {CORC}\textsuperscript{\textregistered} cable with a {C}u former,''
  \emph{Supercond. Sci. Technol.}, vol.~30, no.~8, p. 085012, 2017.

\bibitem{takeuchi2011}
K.~Takeuchi, N.~Amemiya, T.~Nakamura, O.~Maruyama, and T.~Ohkuma, ``Model for
  electromagnetic field analysis of superconducting power transmission cable
  comprising spiraled coated conductors,'' \emph{Supercond. Sci. Technol.},
  vol.~24, no.~8, p. 085014, 2011.

\bibitem{amemiya2013}
N.~Amemiya, R.~Nishino, K.~Takeuchi, M.~Nii, T.~Nakamura, M.~Yagi, and
  T.~Okuma, ``Ac loss analyses of superconducting power transmission cables
  considering their three-dimensional geometries,'' \emph{Physica C:
  Superconductivity}, vol. 484, pp. 148 -- 152, 2013.

\bibitem{tominaka2009}
T.~Tominaka, ``Current and field distributions of a superconducting power
  transmission cable composed of helical tape conductors,'' \emph{Supercond.
  Sci. Technol.}, vol.~22, no.~12, p. 125025, 2009.

\bibitem{stenvall2013}
A.~Stenvall, M.~Siahrang, F.~Grilli, and F.~Sirois, ``Computation of self-field
  hysteresis losses in conductors with helicoidal structure using a 2{D} finite
  element method,'' \emph{Supercond. Sci. Technol.}, vol.~26, no.~4, p. 045011,
  2013.

\bibitem{solovyov2014}
M.~Solovyov, J.~{\v{S}}ouc, and F.~G\"om\"ory, ``{AC} loss properties of
  single-layer {CORC} cables,'' \emph{J. Phys.: Conf. Ser.}, vol. 507, no.~2,
  p. 022034, 2014.

\bibitem{sheng2017}
J.~{Sheng}, M.~{Vojen\v{c}iak}, R.~{Terzio\u{g}lu}, L.~{Frolek}, and
  F.~{G\"om\"ory}, ``Numerical {S}tudy on {M}agnetization {C}haracteristics of
  {S}uperconducting {C}onductor on {R}ound {C}ore {C}ables,'' \emph{IEEE Trans.
  Appl. Supercond.}, vol.~27, no.~4, pp. 1--5, 2017.

\bibitem{fu2018}
S.~{Fu}, M.~{Qiu}, J.~{Zhu}, H.~{Zhang}, J.~{Gong}, X.~{Zhao}, W.~{Yuan}, and
  J.~{Guo}, ``{Numerical Study on AC Loss Properties of HTS Cable Consisting of
  YBCO Coated Conductor for HTS Power Devices},'' \emph{IEEE Trans. Appl.
  Supercond.}, vol.~28, no.~4, pp. 1--5, 2018.

\bibitem{wang2019}
Y.~Wang, M.~Zhang, F.~Grilli, Z.~Zhu, and W.~Yuan, ``{Study of the
  magnetization loss of CORC\textsuperscript{\textregistered} cables using a 3D
  T-A formulation},'' \emph{Supercond. Sci. Technol.}, vol.~32, no.~2, p.
  025003, 2019.

\bibitem{yokoyama2017}
S.~{Yokoyama}, J.~{Lee}, T.~{Imura}, T.~{Matsuda}, R.~{Eguchi}, T.~{Inoue},
  T.~{Nagahiro}, H.~{Tanabe}, S.~{Sato}, A.~{Daikoku}, T.~{Nakamura},
  Y.~{Shirai}, D.~{Miyagi}, and M.~{Tsuda}, ``{Research and Development of the
  High Stable Magnetic Field ReBCO Coil System Fundamental Technology for
  MRI},'' \emph{IEEE Trans. Appl. Supercond.}, vol.~27, no.~4, pp. 1--4, 2017.

\bibitem{yachida2017}
T.~{Yachida}, M.~{Yoshikawa}, Y.~{Shirai}, T.~{Matsuda}, and S.~{Yokoyama},
  ``{Magnetic Field Stability Control of HTS-MRI Magnet by Use of Highly
  Stabilized Power Supply},'' \emph{IEEE Trans. Appl. Supercond.}, vol.~27,
  no.~4, pp. 1--5, 2017.

\bibitem{kajikawa2016}
K.~Kajikawa, S.~Awaji, and K.~Watanabe, ``\BIBforeignlanguage{English}{Ac loss
  evaluation of an {HTS} insert for high field magnet cooled by cryocoolers},''
  \emph{\BIBforeignlanguage{English}{Cryogenics}}, vol.~80, pp. 215--220, 2016.

\bibitem{zhang2017}
H.~Zhang, M.~Zhang, and W.~Yuan, ``An efficient {3D} finite element method
  model based on the {T}--{A} formulation for superconducting coated
  conductors,'' \emph{Supercond. Sci. Technol.}, vol.~30, no.~2, p. 024005,
  2017.

\bibitem{higashi2019}
Y.~{Higashi}, H.~{Zhang}, and Y.~{Mawatari}, ``{Analysis of Magnetization Loss
  on a Twisted Superconducting Strip in a Constantly Ramped Magnetic Field},''
  \emph{IEEE Trans. Appl. Supercond.}, vol.~29, no.~1, pp. 1--7, 2019.

\bibitem{higashi2018}
Y.~Higashi and Y.~Mawatari, ``Electromagnetic coupling of twisted
  multi-filament superconducting tapes in a ramped magnetic field,''
  \emph{Supercond. Sci. Technol.}, vol.~32, no.~5, p. 055010, 2019.

\bibitem{comsol}
\BIBentryALTinterwordspacing
{COMSOL Multiphysics}\textsuperscript{\textregistered}. [Online]. Available:
  \url{www.comsol.com}
\BIBentrySTDinterwordspacing

\bibitem{myers2019}
C.~S. {Myers}, M.~D. {Sumption}, and E.~W. {Collings}, ``{Magnetization and
  Flux Penetration of YBCO CORC$^{\rm TM}$ Cable Segments at the Injection
  Fields of Accelerator Magnets},'' \emph{IEEE Trans. Appl. Supercond.},
  vol.~29, no.~5, pp. 1--5, 2019.

\bibitem{mawatari2011}
Y.~Mawatari, ``Superconducting tubular wires in transverse magnetic fields,''
  \emph{Phys. Rev. B}, vol.~83, p. 134512, 2011.

\bibitem{brandt1996}
E.~H. Brandt, ``Superconductors of finite thickness in a perpendicular magnetic
  field: {S}trips and slabs,'' \emph{Phys. Rev. B}, vol.~54, pp. 4246--4264,
  1996.

\end{thebibliography}
\end{document}